%% file: __main.tex
\documentclass[%
 reprint,
superscriptaddress,
 amsmath,amssymb,
 aps,
pra,
]{revtex4-2}

\usepackage{graphicx}% Include figure files
\usepackage{dcolumn}% Align table columns on decimal point
\usepackage{bm}% bold math

\usepackage{algpseudocode}

\usepackage{hyperref}% add hypertext capabilities
\usepackage{amsthm}
\usepackage{comment}
\usepackage{enumitem}
\usepackage{mathtools}
\usepackage{natbib}
\usepackage{amsfonts}
\usepackage{amssymb}

\DeclareMathOperator*{\argmax}{arg\,max}
\DeclareMathOperator*{\argmin}{arg\,min}

\usepackage{multirow}
\usepackage{booktabs}

\usepackage{upgreek,amsmath,amssymb,amstext,mathrsfs,mathtools,stmaryrd,amsthm,centernot}

\usepackage{fancyvrb}
\usepackage{adjustbox}
\usepackage{float}
\usepackage{makecell}
\usepackage{nicefrac}

\usepackage{todonotes}

\begin{document}

\title{Fooling the Decoder: An Adversarial Attack on Quantum Error Correction}

\author{Jerome Lenssen}
\email{jerome.lenssen@vtt.fi}
\affiliation{VTT Technical Research Centre of Finland, 02150 Espoo, Finland}

\author{Alexandru Paler}
\email{alexandru.paler@aalto.fi}
\affiliation{Aalto University, 02150 Espoo, Finland}

\begin{abstract}
Neural network decoders are essential for fault-tolerant quantum computation, yet their internal mechanisms remain poorly understood. Leading machine learning decoders use recurrent and transformer models (e.g., AlphaQubit), and reinforcement learning (RL) is increasingly used to train such models. In this work we ask whether an RL surface code decoder, DeepQ, can be made to fail systematically. Our white-box search repeatedly draws candidate syndrome volumes from the unmodified noise model and retains the one the decoder's learned $Q$-function is most pessimistic about. At code distance 5 under phenomenological depolarising noise, it cuts the average logical qubit lifetime from roughly $10^5$ cycles to 60. We quantify what this costs: the retained error histories are individually plausible, but the selected trajectory is roughly a $10^{-18}$ tail event under undisturbed noise, so the effect is bought entirely by the ability to select among physically realisable error configurations. Controls confirm this is not an artefact of a weak target: the decoder resists noise fluctuations, tolerates an MWPM referee, and is fault tolerant at checkable distances. The result is a robustness upper bound for one decoder under one noise model, but to our knowledge the first adversarial attack of a machine learning QEC decoder.
\end{abstract}

\maketitle

\section{Introduction}

Machine learning techniques are rapidly transforming the landscape of quantum error correction (QEC), offering promising avenues for building fault-tolerant quantum computers. Neural network decoders have emerged as a powerful alternative \cite{Varsamopoulos2020}, capable of learning complex error patterns and potentially achieving superior decoding performance~\cite{bausch2024}. 

While these advancements hold immense potential, the inherent complexity and ``black-box'' nature of many neural network models present challenges in understanding their inner workings and ensuring their robustness against unforeseen noise or even adversarial attacks.

Motivated by the recent advancements of reinforcement learning, and by the increasing interest in machine learning for QEC, this work answers the question \emph{Can a machine learning QEC decoder be made to fail systematically, and at what cost?} in order to offer preliminary answers to \emph{How robust are machine learning QEC decoders?}. The robustness of general machine learning models is crucial, and it is surprising that QEC methods have not been analysed from this perspective by now. We argue that it is crucial to not only enhance the performance of machine learning decoders but also to analyze their behavior and identify potential vulnerabilities, ultimately paving the way for their confident deployment in future quantum technologies.

The \emph{reliability} question asks whether a decoder degrades gracefully when the noise it meets departs from the noise it was trained on -- drift, spatial inhomogeneity, miscalibration. The \emph{adversarial} question asks whether there exist error histories, permitted by the physical noise model, on which the decoder fails far more often than its average performance suggests, and how hard such histories are to bring about.

Our robustness analysis (Section~\ref{sec:robustness}) addresses the first and serves here as a \emph{control}, establishing that our target is not simply a fragile decoder; the search of Section~\ref{sec:attack} addresses the second. Demonstrating the adversarial attack does not by itself establish a practical security threat, and Section~\ref{sec:threat_model} states precisely what it would take to realise one.

\subsection{Background}

The development of machine learning QEC decoders, starting from~\cite{torlai2017neural}, encompasses a wide range of architectures~\cite{yan2026rethink, lee2026machine}, including recurrent neural networks~\cite{varbanov2025}, convolutional neural networks~\cite{breuckmann2018scalable}, message passing~\cite{maan2025machine}, and increasingly sophisticated transformer models~\cite{bausch2024}. Moreover, machine learning methods are also used as pre-decoders~\cite{chamberland2026fast} or windowed decoders~\cite{zhang2026learning}.  Herein we will not introduce the surface code and the problem of decoding, and refer the interested reader to~\cite{fowler2012surface, demarti2024decoding, battistel2023real}.

Reinforcement learning plays a vital role in training machine learning models, including decoders. Here we focus on the DeepQ RL decoder~\cite{Sweke2020}, which we have chosen due to its simplicity, which makes it a great starting point for a proof of concept attack.

DeepQ works by framing the error correction process as a (hidden) Markov Decision Process. The decoding \emph{agent} learns to make optimal decoding \emph{actions} (applying a single-qubit Pauli operator, $X$ or $Z$, on a specific qubit in the lattice) based on the observed syndrome. The syndrome, a pattern of measurement outcomes indicating the presence of errors, serves as the (observed) state of the environment. The agent chooses an action and the environment (the simulated surface code -- $\mathcal{E}$ in Fig.~\ref{alg:qvalueattack}) transitions to a new state with an updated syndrome.

The agent receives a \emph{reward} based on whether the action moved closer to a valid, error-free state. A deep neural network approximates the $Q$-function, which estimates the expected future reward for taking a specific action in a given environment state. During training, the agent explores the action space by occasionally taking random actions to discover new strategies (exploration), while also exploiting its current knowledge by choosing the action with the highest estimated Q-value (exploitation). The network's weights are updated based on the Bellman equation, minimizing the difference between the predicted Q-value and the actual reward received plus the discounted maximum future Q-value. Through this iterative process of interacting with the surface code environment and updating its Q-function, the agent learns a policy for applying corrections that maintain the encoded quantum information.

\subsection{Related Work}

Robustness of QEC decoders has been studied almost exclusively as a \emph{reliability} property: decoders are evaluated under noise that is mismatched to their training distribution, drifting in time, or spatially inhomogeneous~\cite{Spitz2018, Fowler2010, tan2026resilience}, and thresholds are reported as averages over the noise ensemble.

To our knowledge, no prior work asks whether a machine learning decoder can identify the error histories it handles worst. The security literature on quantum computing has so far concentrated on the hardware and multi-tenancy layers rather than on the classical decoding software, and the adversarial machine learning literature has not treated QEC decoders as targets. This paper is a first step across that gap.

Adversarial attacks on general RL models have been extensively analyzed in the literature. One of the first attacks on machine learning models~\cite{Goodfellow2014} fools an image classifier by slightly perturbing pixel values in the image, leading to a misclassification of the image. To find vulnerabilities of interest, perturbations to the input are typically bounded. For image classification instances, the attack finds perturbations imperceptible to human eyes but detectable by camera sensors. Due to the sequential nature of reinforcement learning tasks, attacks can also consider the temporal aspect. Several attacks on RL agents have been proposed~\cite{Lin2017, Lütjens2019, Ilahi2022}. For example, Lin et al.~\cite{Lin2017} proposed an attack that perturbs the frames of Atari games such as Pong, at specific points in time, fooling the agent into moving in the wrong direction.

\subsection{Contributions}

In the context of deep RL verification~\cite{landers2023deep}, we care about the safety of the decoder (i.e. nothing bad ever happens). To investigate the safety of the decoder, an \emph{adversarial attack} generates adversarial samples that fool the model to produce incorrect decoding decisions. To this end, \emph{we are searching} for adversarial counterexamples, small perturbations to the inputs, such that the decoding outcomes are unexpectedly wrong. Our contributions are the following:
\begin{itemize}
    \item we state an explicit threat model and scope for adversarial analysis of QEC decoders (Section~\ref{sec:threat_model}), and list assumptions which are realistic today and which are not;
    \item we develop a proof of concept attack by starting from DeepQ, and propose a way for \emph{automatically extracting the attack samples};
    \item we quantify the \emph{cost} of the attack (Section~\ref{sec:attack_cost}), i.e. how improbable the error histories it selects are under the undisturbed noise model;
    \item we propose empiric robustness analysis methods on DeepQ, in order to ensure that the attack is not trivial, and we validate the $Q$-function as a proxy for decoding difficulty (Section~\ref{sec:res_lifetime}).
\end{itemize}

% We are explicit about what this paper does \emph{not} do. It targets a single, deliberately simple RL decoder rather than a state-of-the-art one; it uses phenomenological rather than circuit-level noise; and the headline result is obtained without measurement errors. Section~\ref{sec:limitations} collects these limitations and explains which conclusions they do and do not support.

Our paper is organized as follows. We introduce our attack method in Section~\ref{sec:methods}, where Section~\ref{sec:threat_model} states the threat model, Section~\ref{sec:attack} details the attack algorithm, and Section~\ref{sec:robustness} describes how we verify that the attacked decoder is itself sufficiently robust. The results are discussed in Section~\ref{sec:results} -- the attack's effect on the logical qubit lifetime in Section~\ref{sec:res_lifetime}, its run time in Section~\ref{sec:res_duration}, the cost of the selection budget it requires in Section~\ref{sec:attack_cost}, and how the attack can be used to understand the decoder's learned noise model in Section~\ref{sec:res_noise}. We discuss the limitations of our approach and conclude in Section~\ref{sec:conclusion}.

\section{Methods: Shortening a logical qubit's lifetime}
\label{sec:methods}

Attacks can be categorized according to the level of access the attacker has to the model (i.e. white-box versus black-box attacks) \cite{Ilahi2022}, or the method used to generate the input (e.g. gradient-based or combinatorial approaches)~\cite{Chakraborty2018}. 

We investigate an \textit{adversarial attack} based on DeepQ's learned $Q$-function, an approach, that, to our knowledge, has not yet been studied in the context of QEC decoding. Incorporating such adversarial sequences into the replay buffer during training could improve the robustness of the decoding model, or they could be stored in a separate lookup table decoder to support DeepQ in the decoding process~\cite{Das2021}. 

As mentioned, the decoding problem is treated similar to a game whose purpose is to erase errors captured by a \emph{syndrome volume}. The latter is a time-ordered list of error detection events generated by the repeated measurement of the code's stabilizers. The measurements are obtained by treating the surface code protected logical qubit as a black-box (one does not know anything about the logical qubit's state except the measurement result of the code's stabilizers).

Nevertheless, we are treating the decoder as a white-box, while keeping the quantum system as a black-box. As a result, we are presenting a white-box attack. Our attack will have access to all the necessary model parameters to generate valuable adversarial error sequences. Since the syndrome volume is a binary tensor, gradient-based attacks, such as FGSM~\cite{Goodfellow2014}, would require rounding, which reduces its precision. Instead, we consider combinatorial approaches. We are interested in syndrome sequences that have a high chance of occurring in reality; therefore, the attack is based on the noise model of interest -- we sample syndrome volumes from the (black-box) logical qubit using realistic noise parameters (e.g. at a physical error rate of 0.1\% for single qubit gates in the depolarising noise model).

\begin{figure}[t]
\begin{algorithmic}[1]
    \Require Environment $\mathcal{E}$, Action-value function $Q$, $N > 0$
    \State Initialize $T \gets \{\}$
    \While{episode not over}
        \State Sample $N$ syndrome volumes $\{\sigma(E_j)\}_{j \in [1,N]}$ from $\mathcal{E}$
        \State Compute $j^{*} \gets \argmin_{j \in [1,N]} \max_{a \in \mathcal{A}} Q(\sigma(E_j), a)$
        \State Apply the error $E_{j^{*}}$ to hidden state of environment $\mathcal{E}$
        \State Apply associated greedy action $a_{j^{*}}$ to hidden state of environment $\mathcal{E}$
        \State Update the error chain $T \gets T \cup \{E_{j^{*}}\}$
    \EndWhile
    \State \Return $T$
\end{algorithmic}

\caption{$Q$-value informed adversarial attack. We assume a surface code memory experiment is the RL environment $\mathcal{E}$, and a trained decoder for which the Q-value function has been already obtained. The attack samples $N$ syndrome volumes at each step in the episode and chooses the volume resulting in the smallest $Q$-value, thus generating an adversarial error chain $T$. $\sigma(E)$ corresponds to the binary syndrome vector measured by combining the Pauli errors $E$ with the hidden state.}
\label{alg:qvalueattack}
\end{figure}

\subsection{Threat Model and Scope}
\label{sec:threat_model}

Because the term \emph{adversarial attack} carries strong connotations, what follows is best read as a \emph{white-box, offline stress test}: a search for the error histories that a trained decoder handles worst, conducted within the support of a realistic physical noise model. We do not claim a deployable exploit against a fielded quantum computer. The analysis assumes:
\begin{description}
\item[(A1)] \emph{Read access to the trained model.} The analyst knows the decoder's architecture and weights and can evaluate $Q(s,a)$ offline. This is the standard white-box setting of the adversarial machine learning literature~\cite{Ilahi2022}. It is plausible for a decoder distributed as a pre-trained artefact or supplied by a third party, and implausible where the weights remain secret to the operator.

\item[(A2)] \emph{Influence over which physical errors are realised.} Line~5 of Algorithm~\ref{alg:qvalueattack} applies the selected Pauli error $E_{j^{*}}$ to the hidden state of the code. Our procedure therefore models an entity that \emph{selects} among several physically plausible error configurations, rather than one that flips syndrome bits in transit between the device and the decoder. This is a strong assumption and its role is to let us characterise the decoder's worst case, not to describe a realisable intrusion.

\item[(A3)] \emph{A sampling budget, not a physical capability.} At each round, $N$ candidate syndrome volumes are drawn from the \emph{unmodified} noise model and the least favourable is retained. $N$ is thus a search budget. Every individual error configuration we apply is one the device could have produced on its own; what the procedure supplies is selection among them. Section~\ref{sec:attack_cost} quantifies how improbable the resulting history is, and we regard that number, rather than the lifetime reduction alone, as a more exact measure of the threat.

\item[(A4)] \emph{Offline timing.} The search runs offline, at seconds per episode (Section~\ref{sec:res_duration}), which is six or more orders of magnitude slower than the microsecond-scale decoding cycle of a fault-tolerant controller. It could not be executed inline against a running device.

\item[(A5)] \emph{One target, one noise model.} All attack results are for the DeepQ decoder~\cite{Sweke2020} at code distance 5 under phenomenological depolarising noise, without measurement errors.
\end{description}

Under (A1)--(A5), the appropriate reading of our results is a \emph{robustness upper bound}: they measure how badly a trained RL decoder can be made to perform by an entity that knows it completely and may search freely, and they say nothing directly about how such a decoder behaves against an adversary constrained to realistic interfaces.

% We consider this a worthwhile quantity to establish first. A bound of this kind tells the community whether the tail is worth worrying about at all before anyone invests in modelling restricted adversaries, and our answer -- that 
Under these assumptions, we show that there exist decoders, with a perfectly ordinary threshold, that can be driven to fail three orders of magnitude sooner. Tightening (A1) to black-box access and (A2) to syndrome-level perturbation are, in our view, the two most important next steps.%, and we return to them in Section~\ref{sec:limitations}.

\subsection{Adversarial Attack on RL Decoders}
\label{sec:attack}

The herein proposed attack is a $Q$-value informed attack, such that it is based on the $Q$-function of the decoder, which is defined as the expected discounted reward by starting in state $s$ and taking action $a$, following the approach described in~\cite{Lütjens2019}.

If we assume that the decoder's $Q$-function is a good approximation of the real $Q$-function, then, it reflects how pessimistic or optimistic the decoder is about a state $s^{\prime}$. Furthermore, it is known that the agent will act greedily during the evaluation, selecting $a^{*} = \argmax_{a \in \mathcal{A}} Q(s^{\prime}, a)$. From these considerations, a simple algorithm can be derived, forcing the decoder to decode the states about which it is most pessimistic. The pseudocode is in Algorithm~\ref{alg:qvalueattack}.

To evaluate our adversarial attack, we choose the distance 5 DeepQ decoder. The syndrome volumes are sampled from the DeepQ environment with phenomenological depolarizing noise at a physical error rate of $0.001$. At this error rate, the decoder can preserve the code state for up to $10^5$ measurement cycles (middle column of Figure~\ref{fig:robustness_conc_quad_cross} from Appendix).

Our goal is to significantly reduce the qubit lifetime with our method. We evaluate four different approaches to select the syndrome volumes in our attack:
\begin{enumerate}
    \item \textsc{MIN}: Select syndrome volume resulting in smallest $Q$-value.
    \item \textsc{MAX}: Select syndrome volume resulting in largest $Q$-value.
    \item \textsc{MIN-VAR}: Select syndrome volume resulting in smallest variance in $Q$-values of action-state pairs.
    \item \textsc{MAX-VAR}: Select syndrome volume resulting in largest variance in $Q$-values of action-state pairs.
\end{enumerate}

The \textsc{MIN} method is the natural approach, selecting the syndrome volume that the decoder is most pessimistic about decoding. With this method, we expect significantly reduced qubit lifetimes. In contrast, with the \textsc{MAX} approach, we expect the opposite outcome, as larger $Q$-values theoretically correspond to a higher expected reward. Although impractical for an actual attack, it helps to validate our intuition about the $Q$-function.

The other two approaches (\textsc{MIN-VAR} and \textsc{MAX-VAR}) are heuristics. If the $Q$-values have a small variance, it might indicate that the decoder is indifferent about the actions implying that it does not know which action to choose. For large variances, there might be a bias towards certain actions.

\subsection{Robustness Guarantees}
\label{sec:robustness}

\begin{figure}[!ht]
    \centering
    \includegraphics[width=\linewidth]{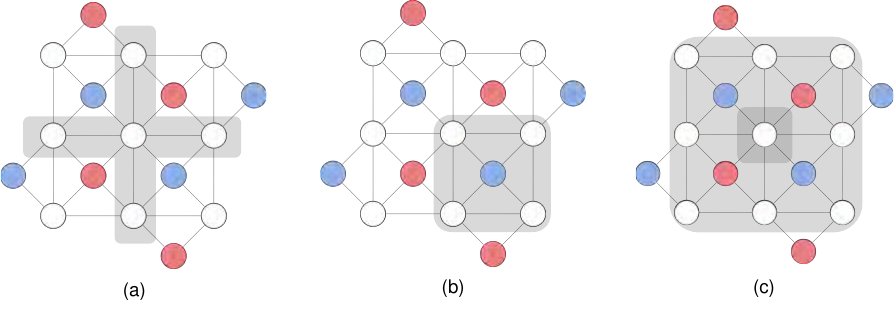}
    \caption[Spatially inhomogeneous Pauli noise]{Spatially inhomogeneous Pauli noise concentrated in specific areas of the surface code. The gray areas mark locations where the qubit error probability is higher. \textbf{(a)} cross noise, \textbf{(b)} quadrant noise, and \textbf{(c)} concentric noise.}
    \label{fig:spatially_inhomogeneous_noise_patterns}
\end{figure}

The attack from the previous section should not be the result of a low quality decoder. Thus, we have to ensure that our RL decoders are robust: a) ensure that the used RL training procedure is stable even if some of the parameters are changed (Section~\ref{sec:role_of_referee_decoder}), b) have a threshold and are not susceptible to noise fluctuations (Section~\ref{sec:exp_robustness_inhomogeneous_time_dependent_noise}, c) the low-distance ones (e.g., 3 and 5) can correct all single- and two-qubit errors (Section~\ref{sec:verify_fault_tolerance_of_decoder}). In the following we detail the robustness analysis methods and the results are in the Appendix.

Regarding noise fluctuations, metrics related to qubit decoherence, such as $T_1$ and $T_2$ times, tend to fluctuate over time, due to environmental noise, requiring continuous device calibration to ensure functionality~\cite{Burnett2019}. Furthermore, qubits on the device are not identical and require individual fine-tuning~\cite{Google2022, Krinner2022}. As a result, decoders must be capable of handling time-dependent variations in noise and non-uniformly distributed error probabilities. The DeepQ decoder performs well for single-qubit depolarizing noise and bit-flip noise at different physical error rates~\cite{Sweke2020}. Other noise models have not been tested so far. 

We evaluate DeepQ in two different scenarios to verify whether they are robust to an environment as described above. Although the proposed noise models are unlikely to be encountered on real devices, they allow us to study whether DeepQ's strategy is flexible enough to cope with diverse noise. For neural network-based image classifiers, it has been demonstrated, that single-pixel changes can lead to significantly worse classification~\cite{Goodfellow2014}. The noise patterns, although not an exhaustive search, give a first indication whether there might be regions of the surface where DeepQ is vulnerable.

\textbf{Spatially inhomogeneous noise}\quad We evaluate DeepQ for noise models where qubits have different error probabilities. In the simplest case, the physical error probabilities are drawn from a normal distribution $\mathcal{N}(p_{\textrm{phys}}, \sigma^2)$, while keeping the measurement error constant at $p_{\textrm{meas}}$.

This approach is evaluated for depolarizing, bit-flip (only $X$ errors) and phase-flip (only $Z$ errors) noise. Furthermore, we evaluate noise ``patterns'', where the noise is concentrated in a specific region of the code. The patterns, referred to as cross, quadrant, and concentric noise, are shown in Figure~\ref{fig:spatially_inhomogeneous_noise_patterns}. For each pattern, depolarizing noise with a physical error rate $p_{\textrm{phys}}$ is used. For cross and quadrant noise, the error probability is increased by $\beta$ in the areas highlighted in gray in Figure~\ref{fig:spatially_inhomogeneous_noise_patterns}. For concentric noise, the qubits at the edge of the surface code are assigned a lower error probability than those in the middle. The error probabilities increase towards the center of the code and lie in the interval $[\nicefrac{p_{\textrm{phys}}}{2}, p_{\textrm{phys}} + \beta]$.

\textbf{Time-dependent noise}\quad For time-dependent Pauli noise, an adaptive edge weight estimator for MWPM has been developed in~\cite{Spitz2018}, increasing the decoding performance compared to a MWPM decoder with fixed weights. We investigate whether DeepQ is able to cope with changing noise by making the physical error rate time dependent
\begin{equation}
    \hat{p}_t = p_{\textrm{phys}} + \beta \sin(2\pi \times \omega t + r_i),
\end{equation} 
where $\beta$ is the same for all qubits and $t$ is the total number of steps in the environment (ignoring resets). In addition, we evaluate qubit-dependent randomized phase offsets $r_i$, such that the qubit error probabilities differ and do not increase at the same time.

We report the average qubit lifetime for each noise model and say that the model is robust if the attained lifetime does not deviate significantly when the bias $\beta$ is increased.

% To be clear about the logic: robustness in this sense is a \emph{precondition} for our attack result to be meaningful, since it rules out the possibility that any perturbation whatsoever degrades the decoder. It is not itself evidence of a vulnerability, and the two sets of experiments should not be read as addressing the same question. 

\section{Results}
\label{sec:results}

In this section, we evaluate the effectiveness of our white-box attack and select the distance 5 DeepQ decoder as a target.

\subsection{Attacking the Lifetime of the Qubit}
\label{sec:res_lifetime}

\begin{figure}[!ht]
    \centering
    \includegraphics[width=0.9\linewidth]{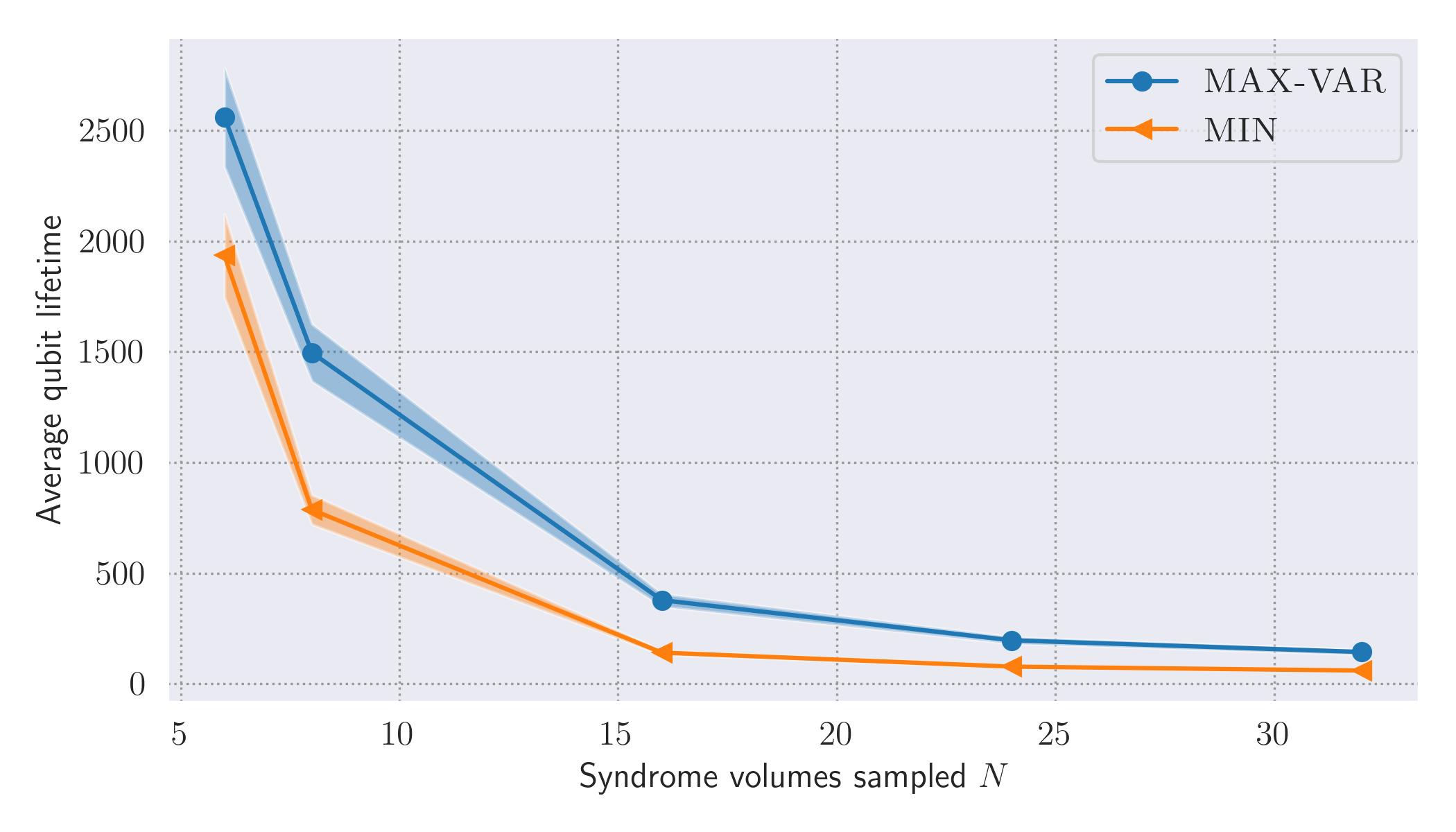}
    \caption[Comparison of adversarial attack syndrome volume selection methods]{Comparison of \textsc{MIN} and \textsc{MAX-VAR} syndrome volume selection method for different numbers of sampled syndrome volumes $N$ per round. The more volumes are sampled, the smaller the average qubit lifetime becomes. The \textsc{MIN} outperforms the \textsc{MAX-VAR} approach.}
    \label{fig:min_var_adversarial_attack_evaluation}
\end{figure}

In our first experiment, we evaluate each method and analyze the effect of increasing the number of sampled syndrome volumes $N$ per round. For each method and $N$ the attack is repeated $500$ times. The results are summarized in Figure~\ref{fig:min_var_adversarial_attack_evaluation}. 

The \textsc{MIN} method proved to be most effective, reducing the average qubit lifetime from $10^5$ cycles to 60 for $N=32$, a reduction of more than three orders of magnitude. This means that, on average, a chain of 12 syndrome volumes had to be generated to result in a logical qubit error. This is notable since, at a physical error rate of $0.001$, the syndrome volumes rarely contain more than one error: the collapse is not produced by burying the decoder in errors, but by choosing which sparse errors occur.

There are two caveats. First, no measurement errors were applied in this experiment, so the figure characterises the decoder's response to data qubit errors alone. Second, the reduction is achieved under assumption (A3), i.e. with a selection budget, and Section~\ref{sec:attack_cost} makes the price of that budget explicit.  

The second best performing method was \textsc{MAX-VAR}, which performed slightly worse than \textsc{MIN}. However, both methods achieve an exponential decay of the qubit lifetime as a function of $N$.

We omitted the results for \textsc{MAX} and \textsc{MIN-VAR} since they resulted in large qubit lifetimes which are irrelevant for an attack. While this was expected for the \textsc{MAX} approach, confirming that DeepQ learned a good approximation of the real $Q$-function, it is surprising for \textsc{MIN-VAR} to perform poorly. The results imply that states for which the action-state value pairs have a higher variance, are more difficult for DeepQ to decode that those with lower variance. A conclusive explanation could not be found on the basis of the observed states.

\emph{Is a low $Q$-value a proxy for decoding difficulty?} The \textsc{MAX} selector tests this assumption directly by running the search in reverse. Selecting the syndrome volume the decoder is most \emph{optimistic} about should \emph{extend} the lifetime rather than shorten it. It does: at physical error rate $0.006$, increasing $N$ from 2 to 4 raised the average qubit lifetime from 302 to 4,482 cycles over 500 repetitions. This licenses \textsc{MIN} as an attack and simultaneously confirms that DeepQ has learned a reasonable approximation of the true $Q$-function. \textsc{MAX} is of course the opposite of an attack, and is better read as a way to identify the syndromes a decoder handles best.

% We stress the limit of this evidence. The two-sided monotonicity above establishes a \emph{statistical} correlation between low $Q$-values and short lifetimes; it does not establish that a low-$Q$ state corresponds to an error chain that flips the logical operator, and the two are not the same claim. A state the network is pessimistic about may reflect broad uncertainty over the action space rather than a genuinely logical-error-inducing configuration. Distinguishing these two explanations would require classifying the selected states by whether the residual error chain is homologically nontrivial, which we have not done.

\begin{figure}[!ht]
    \centering
    \includegraphics[width=\linewidth]{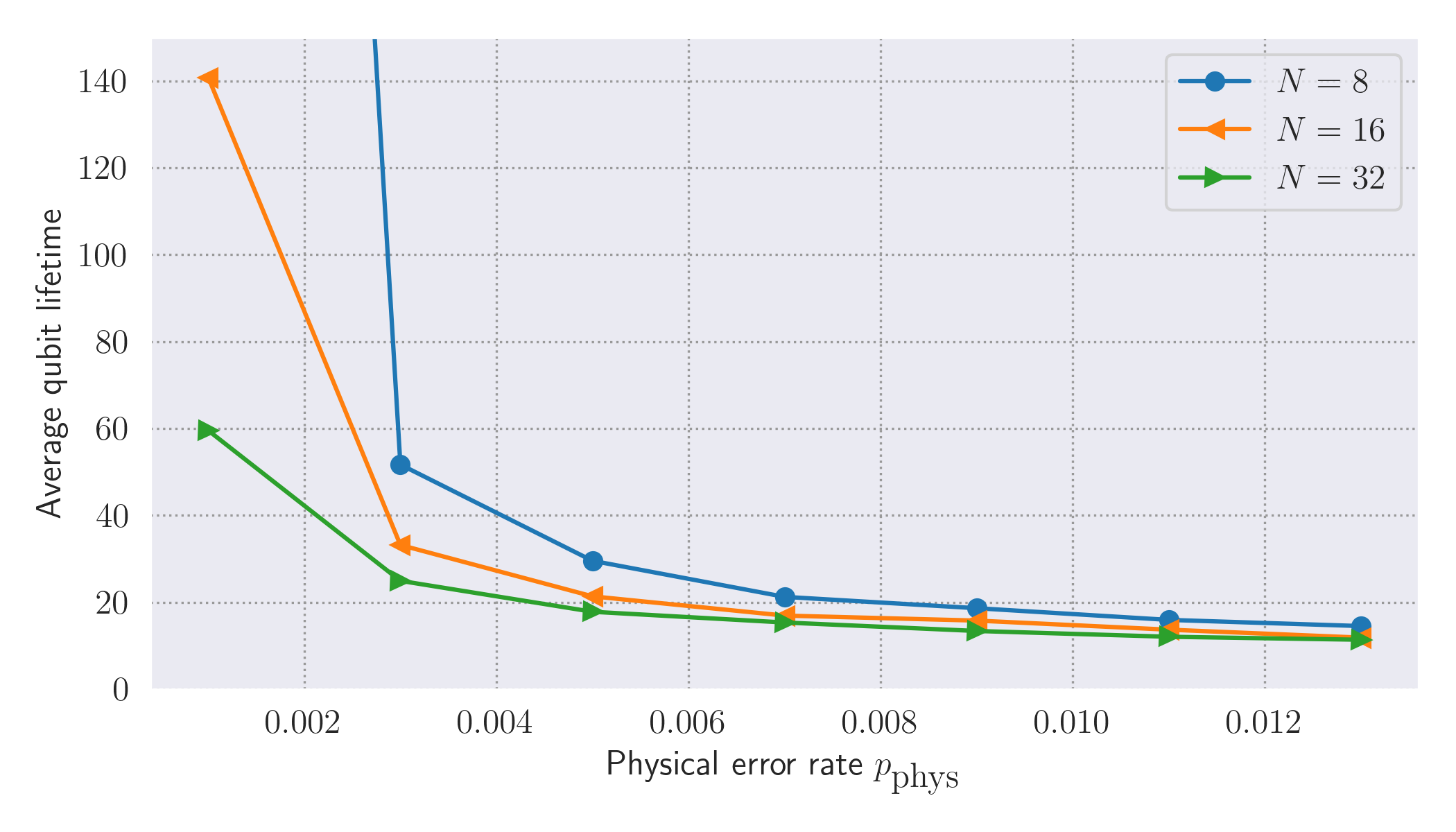}
    \caption[Influence of physical error rate on adversarial attack]{Influence of physical error rate on adversarial attack for $N \in \{8, 16, 32\}$ and \textsc{MIN} syndrome volume selection.}
    \label{fig:adversarial_attack_error_rate_evaluation}
\end{figure}

\subsection{Attack Duration}
\label{sec:res_duration}

We compare the average time needed for an attack based on the \textsc{MIN} method to succeed, for varying $N$. For $N=32$, an attack took on average 1.23 seconds. Halving the number of volumes increased the average to 1.5 seconds and for $N=8$, the time increased to 3.2 seconds per attack. For $N \ge 32$, the syndrome volume generation dominates the run-time, slowing down the attack, however the average qubit lifetime is further reduced. We believe that a sweet-spot exists for $N \in [16,32]$, where the attack is still fast, but the risk of sampling outliers, i.e. syndrome volumes that are difficult to decode but appear seldom in real training, is reduced.

We believe that the run time of our attack can be significantly improved with better compute hardware. Additionally, the number of actors executing the attack could be increased if more samples are needed.

% These timings should nonetheless be put in perspective, since they bound what the method could ever be used for. A fault-tolerant controller must decode a syndrome round in roughly a microsecond, whereas our search requires on the order of a second per episode, i.e. it is around six orders of magnitude too slow. Parallelism and better hardware do not close a gap of that size, and the reason is structural rather than technological: the search must sample and evaluate many candidate futures per round, work that no amount of engineering removes. Our procedure is thus an offline analysis and certification tool, appropriate for auditing a trained decoder before deployment, and not a mechanism that could be operated inline against a running device. This is a limitation of the method as a threat, and a convenience for its intended use.

In Figure~\ref{fig:adversarial_attack_error_rate_evaluation}, we studied the influence of the physical error rate on our attack. As expected, $N$ can be decreased for higher error rates, achieving similar qubit lifetimes.

\subsection{The Cost of the Attack}
\label{sec:attack_cost}

A lifetime reduction is only alarming in proportion to how easy it is to induce, so we now quantify the price of the selection budget in assumption (A3). Two figures matter, and they point in opposite directions.

The first is the sampling cost, which is modest. A successful \textsc{MIN} attack at $N=32$ consumes on average $32 \times 12 \approx 384$ candidate syndrome volumes, of which all but 12 are discarded. Every retained configuration is an ordinary sample from the unmodified depolarising noise model at $p = 0.001$. In this sense the attack is cheap, and the individual error configurations it uses are exactly as plausible as the noise model says they are.

The second is the improbability of the resulting \emph{trajectory}, which is severe. Retaining the worst of $N$ independent draws at each of $L$ successive rounds selects a history that occupies roughly the $N^{-L}$ tail of the undisturbed distribution over histories. For $N = 32$ and $L = 12$,
\begin{equation}
    N^{-L} = 32^{-12} \approx 10^{-18},
\end{equation}
whereas the undisturbed decoder fails spontaneously at a rate near $10^{-5}$ per cycle. The search therefore buys on the order of thirteen orders of magnitude in failure probability, and it buys all of it through selection. This confirms that the vulnerability is real in the sense that such histories \emph{exist} within the support of a realistic noise model, and are findable using nothing but the decoder's own $Q$-function. It equally confirms that the histories are very rare to arise from noise drift, miscalibration, or any of the reliability mechanisms surveyed in Section~\ref{sec:robustness}. An entity able to induce a $10^{-18}$ event on demand is, by construction, not a passive one.

We conclude that we have measured is the decoder's worst case and the selection power required to reach it, not the existence of a practical attacker. Whether any realistic interface to a quantum control stack offers even a small fraction of that selection power is an open question. Our results show that there is a gap between a decoder's average behaviour and its worst case. This gap is very large, and therefore that average-case thresholds alone are an incomplete way to certify a machine learning decoder.

\subsection{Inferring a Noise Model for the Attack}
\label{sec:res_noise}

Finally, we study the error distribution generated by our attack to investigate whether it is possible to infer a noise model from the histogram against which DeepQ would be vulnerable. Herein we want to model the noise statistics that affect the decoder, while in Section~\ref{sec:robustness} we developed simplistic noise models without a priori knowledge stemming from the attack.

To measure the distribution of errors, we run our attack 1,000 times with $N=50$, and the additional constraint that at most 2 errors can be applied during each round. This analysis is similar to the discussion from Section~\ref{sec:robustness}. The constraint is enforced by discarding syndrome volumes which introduce more than two distinct errors to the hidden state. For every discarded volume, a new syndrome volume was sampled until a set of $N$ volumes is obtained. While not a realistic constraint, we hope that by restricting the number of errors, we can better identify regions in which DeepQ is vulnerable.

\begin{figure}[!t]
    \centering
    \includegraphics[width=0.9\linewidth]{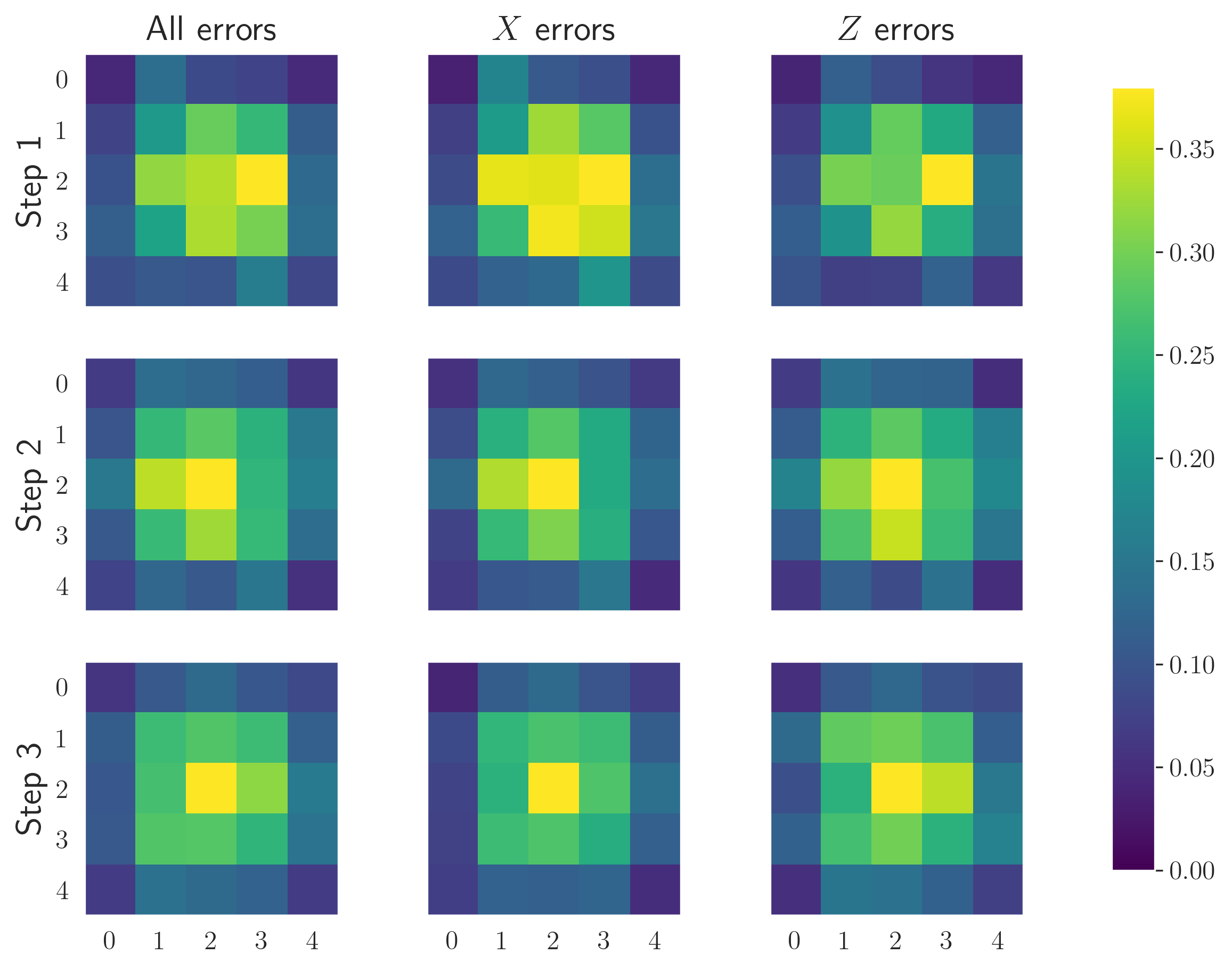}
    \caption[Adversarial attack stepwise Pauli qubit error distribution]{Pauli qubit error distribution for the first three steps of our adversarial attack with depolarizing noise at code distance 5. Each square represents a data qubit. At each step, at most two Pauli errors can occur. The distribution was obtained by averaging the errors obtained by repeating the attack 1,000 times.}
    \label{fig:adversarial_attack_error_distribution}
\end{figure}

Figure~\ref{fig:adversarial_attack_error_distribution} shows the error distribution for the first 3 steps. The errors are concentrated around the center of the surface code, with qubit (2,3) having the highest error probability. In the subsequent steps, the errors are more uniformly distributed around the central qubit. The noise model resembles the concentric noise that we developed in Section~\ref{sec:robustness}. While it was effective, a significant bias was required to decrease the average qubit lifetime of DeepQ.

Therefore, simply averaging the Pauli errors of our attack is insufficient to generate an effective noise model. It could be that the noise model requires to capture temporal transition probabilities from one syndrome volume to another to be effective. Further research is required to provide a conclusive answer to this question.

% \section{Limitations}
% \label{sec:limitations}

\section{Conclusion}
\label{sec:conclusion}

We presented what is, to our knowledge, the first adversarial attack of a machine learning, reinforcement-learning based quantum error correction decoder. Using only the decoder's own learned $Q$-function, our search identifies error histories on which DeepQ fails after roughly 60 measurement cycles rather than the $10^5$ it sustains under undisturbed noise, and the robustness analysis of Section~\ref{sec:robustness} and the Appendix establish that this is not simply the behaviour of a poorly trained decoder.

% We have been deliberate about the strength of the resulting claim. The error configurations the search applies are ordinary samples from a realistic noise model, but the trajectory it assembles is roughly a $10^{-18}$ tail event, and the capability required to induce it on demand is one that no present-day control stack is known to provide. What we have measured, then, is a decoder's worst case and the selection power needed to reach it, rather than a practical exploit. We consider that a useful thing to have measured: it shows that the gap between average-case and worst-case behaviour of a learned decoder is very large, and therefore that reporting thresholds and average lifetimes is an incomplete way to certify one. Whether the gap narrows for stronger decoders, under circuit-level noise, or against an adversary restricted to the syndrome interface are the questions we would most like to see answered next, and Section~\ref{sec:limitations} is explicit about which of them our evidence leaves open. 

We hope this work is reads as an argument that adversarial evaluation belongs in the standard toolkit for validating them. However, we collect here the limitations of our approach, in order to highlight the many ways in which future work can be performed.

\emph{Code capacity, not circuit-level, noise.} All experiments use code capacity noise, practically phenomenological noise but without looking at adversarial measurement errors, which is a property of the DeepQ environment~\cite{Sweke2020} rather than a choice we made freely. Circuit-level noise introduces correlated errors and gate-induced propagation that a phenomenological model does not represent, and these could either amplify the vulnerability, by giving the search richer structure to exploit, or suppress it, by producing syndromes further from those DeepQ was trained on.

% \todo[inline]{more ml decoders referenced}
\emph{A single, deliberately simple target.} We attack only DeepQ, chosen for its simplicity and the availability of documentation sufficient for white-box analysis. It is not a state-of-the-art decoder, and its threshold is modest. It is therefore possible that the vulnerability we report reflects the limits of a small $Q$-network rather than a property of learned decoders in general. We note, against it, that our attack targets the code distance 5 decoder, which does attain near-optimal $p^3$ scaling in the fault tolerance checks of Appendix~\ref{sec:verify_fault_tolerance_of_decoder}; the shortfall in asymptotic scaling that we observe applies to distance 7, not to the decoder actually attacked. The collapse is thus not straightforwardly attributable to a decoder that fails to use its code distance. Establishing whether the effect generalises nonetheless requires repeating the analysis on a modern recurrent or transformer decoder.

\emph{No defence is evaluated.} We suggest in Section~\ref{sec:methods} that adversarial samples could be added to the replay buffer during training, or held in an auxiliary lookup table decoder~\cite{Das2021}, but we have not tested whether either measurably improves robustness. The question of whether these vulnerabilities are removable by adversarial training, or are intrinsic to the function class, is open.

\emph{The attack does not perturb syndrome bits in transit.} As stated in Section~\ref{sec:threat_model}, assumption (A2) grants the ability to select which physically plausible error configuration is realised. We are not aware of a component of a present-day control stack that provides this. Reformulating the search to perturb only the syndrome bits in transit, would change the problem substantially and is left to future work.

\begin{acknowledgements}
We thank Markus M\"uller and Joost-Pieter Katoen for insightful discussions and feedback on the methods. This research was developed in part with funding from the Defense Advanced Research Projects Agency [under the Quantum Benchmarking (QB) program under award no. HR00112230006 and HR001121S0026 contracts]. The views, opinions and/or findings expressed are those of the author(s) and should not be interpreted as representing the official views or policies of the Department of Defense or the U.S. Government.
\end{acknowledgements}

\bibliographystyle{apsrev4-1}
\bibliography{__main}

\input{appendix.tex}

\end{document}

%% file: appendix.tex
\section{Appendix}

\subsection{The Role of the Referee Decoder}
\label{sec:role_of_referee_decoder}

The referee decoder is responsible for terminating episodes during training when it fails to decode the logical qubit state after DeepQ's corrections. In \cite{Sweke2020}, this referee was implemented as a feedforward neural network, requiring architectural adjustments and retraining for each code distance $d$ and noise model. The specific training procedures for the referee, such as batch size or learning rate, were not detailed in the original work.

To eliminate the need to retrain the existing neural network referee at each code distance, we replace it with the \textit{ minimum weight perfect matching (MWPM)} decoder \cite{demarti2024decoding}. The MWPM decoder is easily adaptable to larger code distances, has a polynomial-time complexity, and allows for fine-tuning to a particular noise model via its edge weights.

Since MWPM decodes the errors $X$ and $Z$ separately, we generate two separate matching graphs, one for each type of Pauli error ($Y$ errors are decomposed into $X$ and $Z$). Edge weights are set to 1 by default but are modifiable. For qubits that are stabilized by a single plaquette, we add boundary nodes. The boundary nodes are connected with each other with weight 0. Given the matching graphs, we instantiate two MWPM decoders, one for each Pauli error. Since MWPM predicts corrections on the data qubits, instead of predicting the logical qubit state, we need to modify the check whether an episode ended in the training environment:

\begin{enumerate}
    \item The syndromes are divided into $X$ and $Z$ type error measurements and passed to the corresponding MWPM decoder.
    \item A copy of the hidden state is made and the corrections predicted by the MWPM decoders are applied to the copy.
    \item It is checked whether the logical observable has been flipped. If the state was flipped, we end the episode. Otherwise, the episode continues.
\end{enumerate}

\begin{figure}[ht!]
    \centering
    \includegraphics[width=\linewidth]{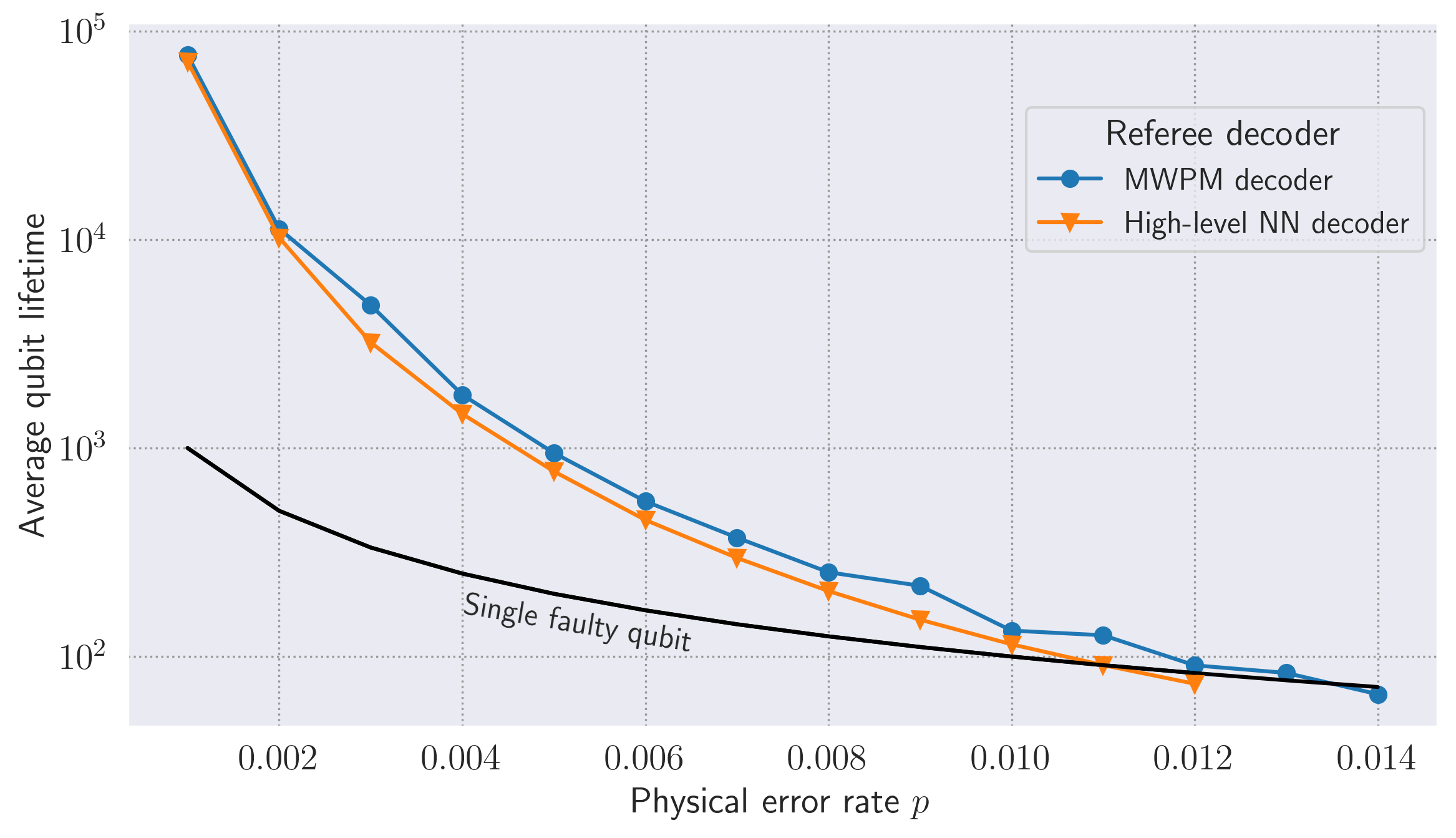}
    \caption[Comparison of MWPM and NN referee decoder for distance 5]{Comparison of MWPM and NN referee decoder for code distance 5 under depolarizing noise. The black line indicates the average qubit lifetime of a single uncorrected qubit.}
    \label{fig:mwpm_nn_referee_comparison}
\end{figure}

We train DeepQ with our new MWPM referee decoder for code distance 5, and compare the resulting performance with the original DeepQ decoder to study whether different referees result in different decoding performances (Figure~\ref{fig:mwpm_nn_referee_comparison}).

Next, we investigate the influence of the referee on DeepQ's training. As it is responsible for terminating episodes, it determines which syndrome patterns DeepQ encounters during training. A referee decoder with worse decoding performance might result in shorter episodes, thereby affecting learning. To test this hypothesis, we intentionally miscalibrate the MWPM edge weights to obtain worse referee decoders for phenomenological depolarizing noise, by randomly drawing weights from a normal distribution centered around 1. Fowler et al.~\cite{Fowler2010} showed that suboptimal edge weights for MWPM lead to poorer decoding performance. We then train DeepQ with different referees, obtained by varying the variance, and compare the resulting average qubit lifetimes.

\begin{figure}[!t]
    \centering
    \includegraphics[width=\linewidth]{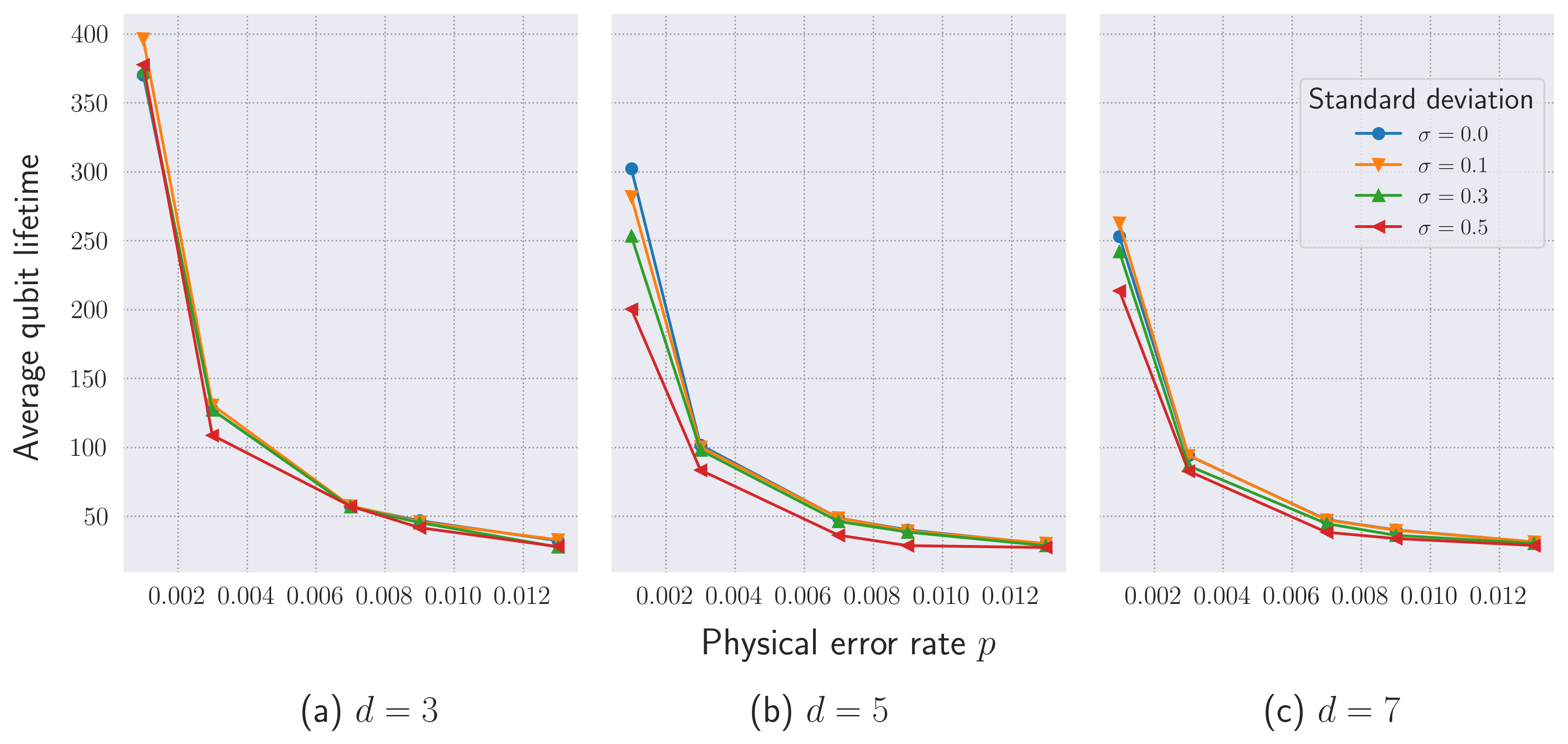}
    \caption[Average qubit lifetime for miscalibrated MWPM weights]{Average qubit lifetime for miscalibrated MWPM referee decoder weights for code distances $3$ (a), $5$ (b), and $7$ (c). MWPM weights are sampled from the normal distribution $\mathcal{N}(1, \sigma^2)$. For each error rate, 20,000 steps were performed in the environment, resetting the data qubits if the referee fails to decode the logical qubit state.}
    \label{fig:mwpm_miscalibration_initial_test}
\end{figure}

The results shown in Figure~\ref{fig:mwpm_miscalibration_initial_test}, indicate that while an underperforming referee decoder has a detrimental effect on the final performance of the DeepQ decoder, it is robust to slight miscalibrations of the referee.

\subsection{Verifying the Fault-Tolerance of the Decoder}
\label{sec:verify_fault_tolerance_of_decoder}

The new MWPM referee introduced in the previous section enables us to train DeepQ decoders up to code distance 9 and thus to estimate the threshold. The threshold achieved for the DeepQ decoder is small compared to other decoders~\cite{Meinerz2022}. We therefore expect that the logical error rate $P_L$ for DeepQ will not follow $p^{\lfloor\nicefrac{(d+1)}{2}\rfloor}$, for error rates below the threshold, but with a smaller exponent.

In theory, the distance $d$ surface code enables the detection and correction of up to $\lfloor \nicefrac{(d-1)}{2} \rfloor$ simultaneous errors. Therefore, an optimal decoder should not make any decoding errors in this regime, otherwise it does not make efficient use of the information provided by the QEC code. To verify whether DeepQ decodes optimally, we would need to decode all possible error patterns with up to $\lfloor \nicefrac{(d-1)}{2} \rfloor$ simultaneous errors. Since explicit enumeration of all errors is not feasible for larger code distances, we propose two experiments, gathering preliminary results on DeepQ's decoding capabilities.

First, we let DeepQ decode all possible error patterns up to three simultaneous $X$ and $Z$ errors. We report the number of logical errors after one round of decoding and the number of remaining flipped syndromes. To keep the enumeration tractable, we restrict this analysis to data qubit errors and exclude measurement errors; consequently these checks establish that the decoder uses the code's error-correcting structure for physical errors, and they make no claim about its behaviour under measurement noise. The remaining number of error patterns for $n$ simultaneous errors is given by
\begin{equation*}
    \binom{d^2}{n} \times 2^n.
\end{equation*}

% https://tex.stackexchange.com/questions/2441/how-to-add-a-forced-line-break-inside-a-table-cell
\begin{table}[!t]
    \centering
    \tiny
    \begin{tabular}{ccrrrr}
        \toprule
        \makecell{\textbf{Code}\\ \textbf{distance}} &  \makecell{\textbf{Number}\\ \textbf{of errors}} & \makecell{\textbf{Error}\\ \textbf{patterns}} &  \multicolumn{1}{c}{\makecell{\textbf{Failure}\\ \textbf{rate in \%}}} & \makecell{\textbf{Referee required}\\\textbf{for final decoding}} & \makecell{\textbf{Sum of remaining}\\ \textbf{syndromes}} \\
        \midrule
        \multirow{3}{*}{$d=3$} & $1$ & $18$ & 0.0 & 8 & 8\\
        & $2$ & $144$ & 32.64 & 96 & 163\\ 
        & $3$ & $672$ & 53.27 & 527 & 1,116 \\
        \midrule
        \multirow{3}{*}{$d=5$} & $1$ & $50$ & 0.0 & 0 & 0\\
        & $2$ & 1,200 & 0.5 & 1,139 & 1,559 \\
        & $3$ & 18,400 & 4.93 & 17,868 & 47,108\\
        \midrule
        \multirow{3}{*}{$d=7$} & $1$ & $98$ & 0.0 & 21 & 21 \\
        & $2$ & 4,704 & 0.0 & 4,008 & 6,506 \\
        & $3$ & 147,392 & 0.37 & 144,244 & 423,040 \\
        \bottomrule
    \end{tabular}
    \caption[DeepQ decoding performance for simultaneous $X$ and $Z$ errors]{Decoding performance of DeepQ in the presence of $1$, $2$ and $3$ simultaneous $X$ and $Z$ errors. No measurement errors were applied. The number of times the referee decoder was required to decode the final state as well as the sum of remaining flipped syndromes after DeepQ decoding is reported. MWPM was used as referee decoder.}
    \label{tab:influence_error_chains}
\end{table}

The results are summarized in Table~\ref{tab:influence_error_chains}. For code distances $3$ and $5$ we observe that the logical error rate scales nearly proportionally to $p^2$ and $p^3$. As predicted, the exponent is slightly lower than the theoretical optimum. Across all code distances, the decoder trained with an error rate of $0.011$ resulted in the lowest LER. For distance $7$, the decoder did not achieve an asymptotic scaling rate of $p^4$, however, $P_L$ decreases at a faster rate than for distance $5$ which indicates that scaling the code distance is beneficial. We note for later reference that the distance 5 decoder, which is the target of the attack in Section~\ref{sec:res_lifetime}, does attain close to its theoretically expected $p^3$ scaling. The attacked decoder is therefore not one that fails to exploit its code distance, which rules out the most obvious benign explanation for the lifetime collapse we report.

Second, we analyze how the logical error rate scales for physical error rates below the threshold. Fowler et al.~\cite{Fowler2012}, showed that $P_L$ can be empirically approximated with the Ansatz
\begin{equation}
\label{eq:logical_error_rate_ansatz}
    P_L \approx \alpha \times (\nicefrac{p}{p_{\textrm{th}}})^{\left\lfloor\nicefrac{(d+1)}{2}\right\rfloor}.
\end{equation}

The idea behind Equation~\ref{eq:logical_error_rate_ansatz} is that an optimal decoder for an odd code distance $d$ cannot distinguish between error chains of length $d_e = \lfloor\nicefrac{(d+1)}{2}\rfloor$ and shorter chains of length $\lfloor\nicefrac{(d-1)}{2}\rfloor = \lfloor\nicefrac{d}{2}\rfloor$. As shorter chains are more probable, longer chains will be misclassified as shorter chains, resulting in a logical error. 

\begin{figure}[!t]
    \centering
    \includegraphics[width=\linewidth]{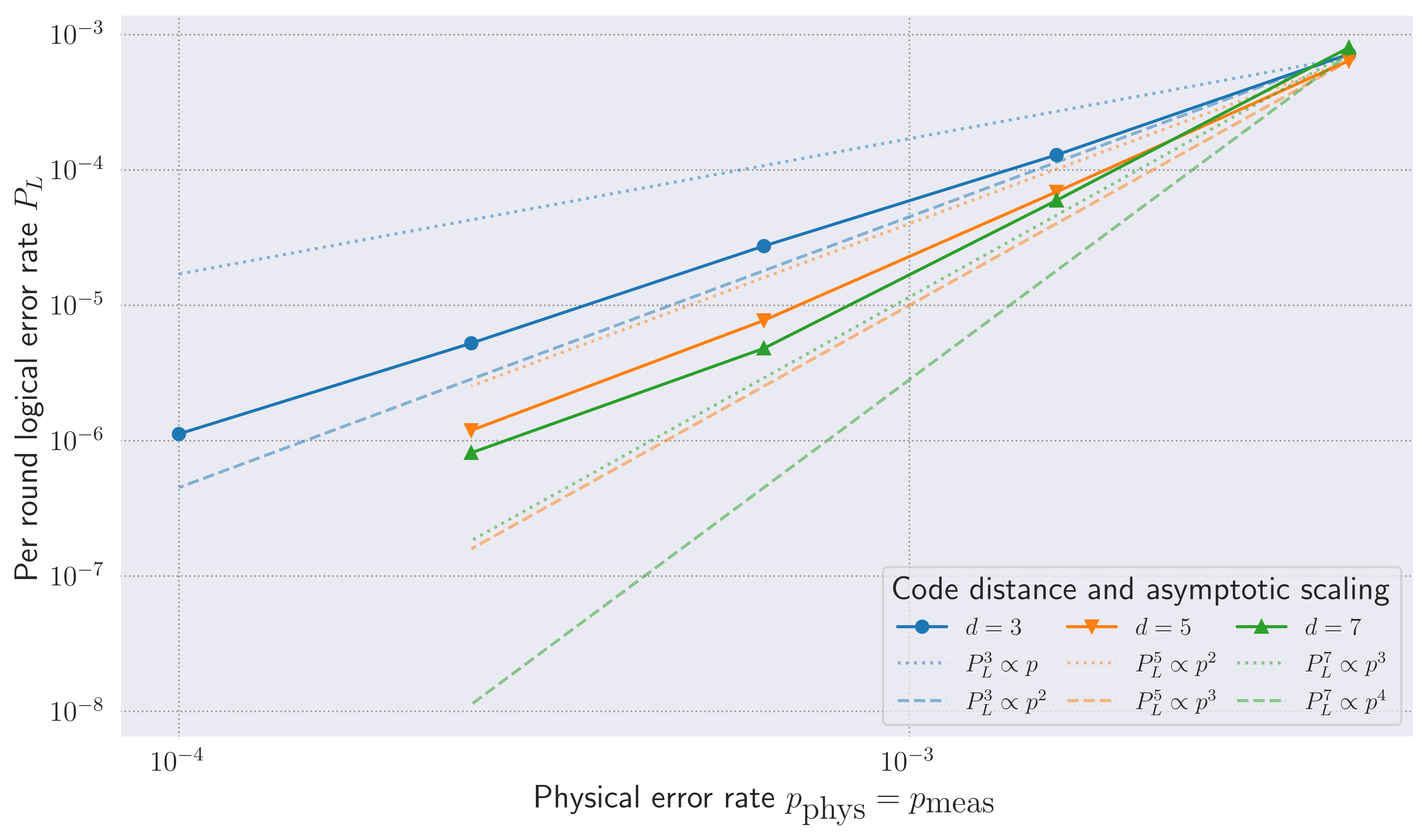}
    \caption[Asymptotic scaling of logical error for DeepQ decoder]{Asymptotic scaling of logical error per round for physical error rates below the threshold $p_{\textrm{th}} \approx 0.004$ of DeepQ.}
    \label{fig:asymptotic_scaling}
\end{figure}

We restricted our estimate of the logical error rate $P_L$ in Figure~\ref{fig:asymptotic_scaling} to $100$ decoding failures.

\subsection{Robustness to Inhomogeneous and Time-dependent Noise}
\label{sec:exp_robustness_inhomogeneous_time_dependent_noise}

Each noise model is tested for physical error rates up to $0.013$, in increments of $0.001$. Each rate is tested for 200 episodes.
After every 10 episodes, we instantiate a new noise model, by sampling new qubit error probabilities according to the noise model's error distribution. We do this to minimize the influence of potential outliers. For the conventional noise models: bit-flip, phase-flip, and depolarizing noise, the standard deviations lie in the range of $\sigma \in [0.0001, 0.0005]$. 

The inhomogeneous depolarizing noise resulted in average qubit lifetimes similar to those of the depolarizing noise model. This was expected since the agent was trained for the latter. The inhomogeneous bit-flip and phase-flip noise resulted in lower average qubit lifetimes. This is explained by the fact that logical error chains form more quickly when restricted to a single Pauli error type.

\begin{figure}[!t]
    \centering
    \includegraphics[width=\linewidth]{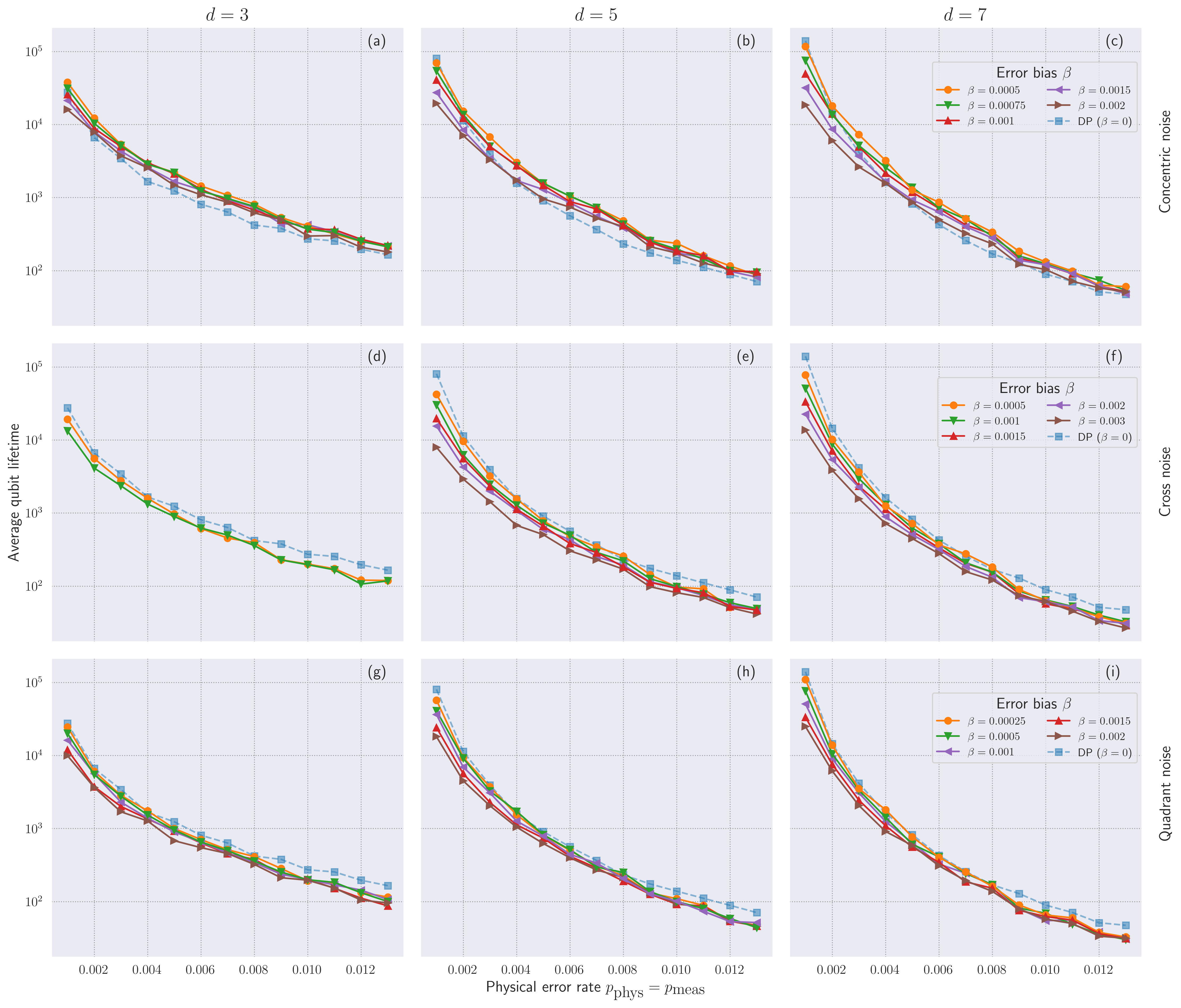}
    \caption[Robustness evaluation for spatially inhomogeneous Pauli noise]{Robustness evaluation for spatially inhomogeneous Pauli noise. Qubit error probabilities are drawn from $\mathcal{N}(p_{\textrm{phys}}, \sigma^2)$. Columns represent code distances $d=3, 5,$ and $7$. Graphs \textbf{(a-c)} concentric noise, \textbf{(d-f)} cross noise, and \textbf{(g-i)} quadrant noise. As a reference, the light-blue, dashed line represents the average qubit lifetime of the best performing agent under depolarizing noise.}
    \label{fig:robustness_conc_quad_cross}
\end{figure}

For noise models with spatially concentrated noise, we choose higher biases, since the results above indicate that DeepQ is insensitive to small fluctuations in error probability. As can be seen in Figure~\ref{fig:robustness_conc_quad_cross}, a higher bias resulted in a noticeably lower average qubit lifetime for small values of $p_{\textrm{phys}}$ for all noise models. The differences are less pronounced at higher error rates, as the added bias $\beta$ is smaller relative to the base physical error rate. Interestingly, for concentric noise, the average qubit lifetime is higher than for depolarizing noise, although the qubit error probabilities at the center are higher. We believe that the noise model makes it less likely for logical error chains to occur since at the edges the error probability decreases. Furthermore, it indicates that DeepQ correctly handles errors in the center, without propagating them to the edges. Finally, for time-dependent noise, similar average qubit lifetimes were observed.

We showed that DeepQ's performance is consistent across a variety of noise models and that the scaling of the average qubit lifetime for inhomogeneous noise follows closely the one for homogeneous depolarizing noise. We conclude that since almost all of the noise models result in the same decrease in average qubit lifetime for higher error rates, DeepQ is relatively robust to fluctuations in noise.